\documentclass[twocolumn,showpacs,preprintnumbers,amsmath,amssymb,prl]{revtex4}

\usepackage{graphicx}
\usepackage{dcolumn}
\usepackage{bm}
\usepackage{amsmath}
\usepackage{amsthm}

\newtheorem{theorem}{Theorem}
\newtheorem{lemma}{Lemma}
\newtheorem{observation}{Observation}
\newtheorem{definition}{Definition}

\newcommand{\ket}[1]{\big|#1\big>}
\newcommand{\bra}[1]{\big<#1\big|}

\newcommand{\proj}[1]{\ket{#1}\bra{#1}}

\newcommand{\ot}[0]{\otimes}
\newcommand{\beq}{\begin{equation}}
\newcommand{\eeq}{\end{equation}}
\newcommand{\tr}[1]{{\rm Tr}(#1)}
\newcommand{\Tr}{{\rm Tr}}
\newcommand{\idmap}{{\rm id}}

\begin{document}


\title{No-local-broadcasting theorem for quantum correlations}

\author{Marco Piani$^1$, Pawel Horodecki$^2$ and Ryszard Horodecki$^3$}
\affiliation{$^1$ Institut f\"ur Theoretische Physik, Universit\"at Innsbruck, Technikerstrasse 25, A-6020 Innsbruck, Austria\\
$^2$ Faculty of Applied Physics and Mathematics, Technical University of Gda\'nsk, 80-952 Gda\'nsk, Poland\\
$^3$ Institute of Theoretical Physics and Astrophysics, University of Gda\'nsk, 80-952 Gda\'nsk, Poland}

\date{\today}

\begin{abstract}
We prove that the correlations present in a multipartite quantum state have an \emph{operational} quantum character as soon as the state does not simply encode a multipartite classical probability distribution, i.e. does not describe the joint state of many classical registers. Even unentangled states may exhibit such \emph{quantumness}, that is pointed out by the new task of \emph{local broadcasting}, i.e. of locally sharing pre-established correlations: this task is feasible if and only if correlations are classical and derive a no-local-broadcasting theorem for quantum correlations. Thus, local broadcasting is able to point out the quantumness of correlations, as standard broadcasting points out the quantum character of single system states. Further, we argue that our theorem implies the standard no-broadcasting theorem for single systems, and that our operative approach leads in a natural way to the definition of measures for quantumness of correlations.
\end{abstract}

\maketitle

The characterization of correlations present in a quantum state has recently drawn much attention~\cite{zurek,HV,compendiumlocal,groisman1,groisman2}. In particular, efforts have been made to analyze whether and how correlations can be understood, quantified and classified as either classical or quantum. Even if such classical/quantum distinction may not be possible in clear-cut terms, understanding to some extent the \emph{quantumness} of correlations is not only relevant from a fundamental point of view, but also in order to make more clear the origin of the quantum advantage~\cite{linden,groisman2}, in fields like quantum computing and quantum information~\cite{NC}, with respect to the classical scenario. Therefore, while entanglement~\cite{reviewent} may be considered \emph{the} essential feature of quantum mechanics, it is relevant to study how and in what sense even correlations present in unentangled states may exhibit a certain quantum character.

In this Letter we provide an \emph{operational characterization} of those multipartite states
whose correlations may be considered as completely classical, hence, by contrast, also of quantumness. We do this in two ways. First, we consider the process of extracting classical correlations (correlations that can be transferred to classical registers) from quantum states, and we prove that this classical correlations amount to the total correlations if and only if the quantum state can be interpreted from the very beginning as a joint probability distribution that describes the state of classical registers. Second, we consider local broadcasting, i.e. the procedure of locally distributing pre-established correlations in order to have more copies of the original state 
\cite{footnote}.
Again, we prove that local broadcasting is feasible if and only if all correlations are classical and the state is just a classical probability distribution. We further generalize the latter approach, showing that what really counts is the amount of correlations, as measured by mutual information.
All the results presented here are valid for the multipartite case, when bipartite mutual information is substituted by one of its most natural multipartite versions. For the sake of clarity, we derive them in the bipartite case.

We start by recalling~\cite{Werner1989,compendiumlocal,groisman2} several definitions that make clear what we mean when we discuss classicality and quantumness, both of bipartite states and of correlations.
\begin{definition}
A bipartite state $\rho$ is: (i) \emph{separable}~\footnote{In the seminal paper by Werner~\cite{Werner1989} the term \emph{classically correlated} is used, but for the sake of clarity we prefer here to say ``separable''.} if it can be written as $\sum_i p_k \sigma_k^A\ot\sigma_k^B$,
where $p_k$ is a probability distribution and each $\sigma_k^X$ is a quantum state, and \emph{entangled} if non-separable; (ii) \emph{classical-quantum} (CQ) if it can be written as
$\sum_i p_i \proj{i}\ot\sigma_i^B$,
where $\{\ket{i}\}$ is an orthonormal set, $\{p_i\}$ is a probability distribution and $\sigma_i^B$ are quantum states; (iii) \emph{classical-classical} (CC), or \emph{(strictly) classically correlated} \cite{compendiumlocal,groisman2}, if there are two orthonormal sets $\{\ket{i}\}$ and $\{\ket{j}\}$ such that $\rho=\sum_{ij}p_{ij}\proj{i}\ot\proj{j}$,
with $p_{ij}$ a joint probability distribution for the indexes $(i,j)$.
\end{definition}
One could consider a CC state to correspond simply to the embedding into the quantum formalism of a classical joint probability distribution.
It is possible to go from a bipartite quantum state to a CQ state and further to a CC state by local measurements.
\begin{definition}
A (quantum-to-classical) measurement map~\footnote{All maps will be understood as channels, i.e. trace preserving and completely positive maps.} $\mathcal{M}$ acts as
$\mathcal{M}[X]=\sum_i\tr{M_iX}\proj{i}$,
where $\{M_i\}$ is a POVM, i.e. $0\leq M_i\leq\openone$ and $\sum_iM_i=\openone$, and $\{\ket{i}\}$ is a set of orthonormal states.
\end{definition}
A measurement map performs a POVM measurement and writes the result in a classical register (i.e., that can be perfectly read and copied), thus any POVM corresponds to a measurement map. Hence, to any bipartite state $\rho$ and any POVM $\{M_i\}$ (on $A$, in this case) we can associate a CQ state
$\rho^{CQ}(\{M_i\})=(\mathcal{M}_A\otimes\idmap_B)[\rho]=\sum_i p_i \proj{i}\ot\sigma^B_i$,
where $\mathcal{M}_A$ is the measurement map associated to the POVM, so that $p_i=\tr{M^A_i \rho}$ and $\sigma^B_i=\Tr_A(M^A_i\rho)/p_i$. Similarly, given POVMs $\{M_i\}$ and $\{N_j\}$ on $A$ and on $B$ respectively, we can associate to $\rho$ the CC state $\rho^{CC}(\{M_i\},\{N_j\})=(\mathcal{M}_A\otimes \mathcal{N}_B)[\rho]=\sum_{ij}p_{ij}\proj{i}\ot\proj{j}$,
with $\mathcal{M}_A$, $\mathcal{N}_B$ the two local measurement maps associated to the two POVMs, and $p_{ij}=\tr{M^A_i\ot N^B_j \rho}$. Notice that that in this case one may always think that the passage from the initial state $\rho$ to the CC state $\rho^{CC}(\{M_i\},\{N_j\})$ happens in two separate (and commuting) steps corresponding to the two local POVMs.

Both from an axiomatic and an operative point of view, we are led to look at Mutual Information (MI) as a \emph{measure of total correlations}~\cite{HV,groisman1}.
\begin{definition}
\emph{(Quantum) mutual information} $I(\rho_{AB})$ of a bipartite quantum state $\rho_{AB}$ is given by $I(\rho_{AB})=S(A)+S(B)-S(AB)$,
where $S(X)=S(\rho_X)=-\tr{\rho_X \log \rho_X}$ is the von Neumann entropy of $\rho_X$.
\end{definition}
Quantum Mutual Information (QMI) is the generalization to the quantum scenario of the classical MI for a joint probability distribution $\{p^{AB}_{ij}\}$: $I(\{p^{AB}_{ij}\})=H(\{p^A_i\})+H(\{p^A_i\})-H(\{p^{AB}_{ij}\})$, with $p^A_i=\sum_j p^{AB}_{ij}$ the marginal distribution for $A$ (similarly for $B$), and $H(\{q_k\})=-\sum_k q_k \log q_k$ is the Shannon entropy of the classical distribution $\{q_k\}$. QMI can be written as the relative entropy between the total bipartite state and the tensor product of its reductions, i.e. $I(\rho_{AB})=S(\rho_{AB}||\rho_A\otimes\rho_B)$, with $\rho_X=Tr_{Y}(\rho_{XY})$. Thus, QMI is positive, and vanishes only for factorized states. Most importantly, it cannot increase under local channels $\Lambda_A\otimes \Gamma_B$, i.e. $I(\rho_{AB})\geq I((\Lambda_A\otimes \Gamma_B)[\rho_{AB}])$~\cite{NC}. 

From an operative point of view, QMI provides the classical capacity of a noisy quantum channel when entanglement is a free unlimited resource~\cite{bennett-cap}. Moreover, for a given state $\rho^{AB}$, $I(\rho^{AB})$ gives the smallest rate of classical randomness necessary and sufficient to  erase all correlations between $A$ and $B$ in the asymptotic setting~\cite{groisman1}.

We will consider two other measures of correlations.
\begin{definition}
Given a bipartite state $\rho_{AB}$ we define:
the \emph{CQ mutual information} as $I_{CQ}(\rho_{AB})=\max_{\{M_i\}}I(\rho^{CQ}(\{M_i\}))$;
the \emph{CC mutual information} as $I_{CC}(\rho_{AB})=\max_{\{M_i\},\{N_j\}}I(\rho^{CC}(\{M_i\},\{N_j\}))$.
\end{definition}
The maxima are taken with respect to (local) measurement maps. Notice that both CQ mutual information and CC mutual information correspond to the QMI of the state after a \emph{local} processing, more precisely after the application of a measurement map. $I_{CC}$ corresponds exactly to the classical MI of the joint classical distribution $p_{ij}=\tr{M_i\otimes N_j \rho}$.
$I_{CQ}$ was considered -- though not in terms of MI -- in~\cite{HV} as a measure of classical correlations, but one may argue that in principle there is still a certain degree of quantumness in the CQ state entering in the corresponding definition. $I_{CC}$ was first defined in~\cite{entpur} and
provides the \emph{maximum amount of the correlations that are present in the state and that can be considered classical, in the sense that can be revealed by means of local measurements, and in this way transfered from the quantum to the classical domain (i.e. recorded in classical registers)}. We have already seen that MI does not increase under local operations. In~\cite{HV} this was proved also for $I_{CQ}$, and the same holds for $I_{CC}$, as local operations on both sides can be absorbed in the measurements. Moreover, $I,I_{CQ},I_{CC}$ are related by local operations themselves and each of them vanish only for uncorrelated state~\cite{HV,locking}. We collect this results in the following
\begin{observation}
\label{obs:ICC}
Mutual information functions $I,I_{CQ},I_{CC}$: (i) are non-increasing under local operations;
(ii) satisfy $I\geq I_{CQ} \geq I_{CC}\geq0$; (iii) vanish if and only if the state is factorized.
\end{observation}

We will prove, with the help of simple lemmas, that all quantum states, that are not CC from the beginning, contain correlations that are not classical, in the sense made precise by Theorem \ref{thm:cc}.
\begin{lemma}
\label{lem:holevo}
Given a CQ state $\rho=\sum_i p_i \proj{i}\ot\sigma_i^B$, we have $I(\rho)=I_{CQ}(\rho)=\chi(\{p_i,\sigma_i\})$, with the Holevo quantity $\chi(\{p_i,\sigma_i\})=S(\sum_i p_i \sigma_i)-\sum_i p_i S(\sigma_i)$. Moreover, we have $I(\rho)=I_{CC}(\rho)$ if and only if the states $\sigma_i^B$ commute and $\rho$ is CC.
\end{lemma}
\begin{proof}
In order to prove $I(\rho)=I_{CQ}(\rho)$, consider the measurement on $A$ corresponding to a complete measurement on the basis comprising the orthogonal states $\{\ket{i}\}$. $I(\rho)=\chi(\{p_i,\sigma_i\})$ is checked straightforwardly. Thus, $I_{CC}(\rho)$ is the classical MI between two parties, where party $A$ sends a state $\sigma_i$ labeled by $i$ with probability $p_i$, and $B$ proceeds to a generalized measurement that gives outputs $j$ with conditional probabilities $p(j|i)$~\cite{NC}. It is known~\cite{holevo} that $\chi$ is an upper bound to the classical MI of $\{p_{ij}=p_ip(j|i)\}$, that is saturated if and only if the states $\sigma_i$ commute, i.e. can be diagonalized in the same basis.
\end{proof}
\begin{lemma}
\label{lem:petz}
If $I((\Lambda_A\ot\Gamma_B)[\rho])=I(\rho)$, there exist $\Lambda_A^*$ and $\Gamma_B^*$ such that $(\Lambda_A^*\ot\Gamma_B^*)\circ (\Lambda_A\ot\Gamma_B)[\rho]=\rho$.
\end{lemma}
\begin{proof}
A theorem~\cite{Petz-equality} by Petz states that, given two states $\rho,\sigma$ and a channel $\Lambda[Y]=\sum_i K_i Y K^\dagger_i$, then $S(\rho||\sigma)=S(\Lambda[\rho]||\Lambda[\sigma])$ if and only if there exists a channel $\Lambda^*$ such that $\Lambda^*[\Lambda[\rho]]=\rho$ and $\Lambda^*[\Lambda[\sigma]]=\sigma$. Moreover, the action of $\Lambda^*$ on $\Lambda[\sigma]$ can be given the explicit expression
$\Lambda^*[X]
=
\sigma^{\frac{1}{2}}
\Lambda^T
\Big[
(\Lambda[\sigma])^{-\frac{1}{2}}
X
(\Lambda[\sigma])^{-\frac{1}{2}}
\Big]
\sigma^{\frac{1}{2}}$, where $\Lambda^T[Y]=\sum_i K^\dagger_i Y K_i$. With this result, if furthermore  $\sigma=\sigma_A\otimes \sigma_B$ and $\Lambda=\Lambda_A\ot\Gamma_B$, one easily checks that $\Lambda^*=\Lambda_A^*\ot\Gamma_B^*$.
\end{proof}
We are now ready to state our first main result.
\begin{theorem}
\label{thm:cc}
We have $I_{CC}(\rho)=I(\rho)$ if and only if $\rho$ is classical-classical.
\end{theorem}
\begin{proof} If the state is CC, it is immediate to check that $I_{CC}=I$. On the other hand, let us assume $I(\rho)=I_{CC}(\rho)=I\big(\rho^{CC}(\{M_i\},\{N_j\})\big)$, with $\rho^{CC}(\{M_i\},\{N_j\})=\sum_{ij}p_{ij}\proj{i}\ot\proj{j}$ for some optimal $\{M_i\},\{N_j\}$. Thanks to Lemma \ref{lem:petz} we have that there exist maps $\mathcal{M}^*$ and $\mathcal{N}^*$ which invert the measurement maps, i.e. such that $\rho=(M^*\ot N^*)[\rho^{CC}]=\sum_{ij}p_{ij}M^*[\proj{i}]\ot N^*[\proj{j}]$. Let us consider
$\tilde{\rho}^{QC}=(M^*\ot \idmap)[\rho^{CC}]
								=\sum_{j}p^B_j\sigma^A_j\ot \proj{j}$,
where $p^B_j=\sum_ip_{ij}$ and $\sigma^A_j=\sum_i p_{ij}/p^B_j M^*[\proj{i}]$. 
This is a QC state such that $I(\tilde{\rho}^{QC})=I_{CC}(\tilde{\rho}^{QC})=I_{CC}(\rho)=I(\rho)$. Therefore, all $\sigma^A_j=\sum_k q^{(j)}_k\proj{\phi_k}$ are diagonal in the same basis $\{\ket{\phi_k}\}$ by Lemma \ref{lem:holevo}. The original state can now be written as
$\rho=\sum_{j}p^B_j\sigma^A_j\ot N^*[\proj{j}]=\sum r_k \proj{\phi_k}\otimes \tau_k$,
where $r_k=\sum_j p^B_j q^{(j)}_k$ and $\tau_k=\sum_j \frac{p^B_j q^{(j)}_k}{r_k} N^*[\proj{j}]$. Thus we have found that $\rho$ is a CQ state with $I=I_{CC}$, therefore it is CC, again by Lemma \ref{lem:holevo}.
\end{proof}

We depict here another operational way to characterize CC states which regards local broadcastability. We first recall the standard broadcasting condition~\cite{barnum-broad}.
\begin{definition}
Given a state $\rho$ we say that $\tilde{\rho}_{XY}$ is a \emph{broadcast state} for $\rho$ if $\tilde{\rho}_{XY}$ satisfies $\tilde{\rho}_{X}=\tilde{\rho}_{Y}=\rho$. 
\end{definition}
We now specialize to the bipartite scenario $\rho=\rho_{AB}$. In this case, one can consider two cuts: one between the copies, and one between the parties. The latter defines locality. Thus, the copies are labeled by $X=AB$ and $Y=A'B'$, while the two parties are $(A,A')$ and $(B,B')$.
\begin{definition}
We say that the state $\rho=\rho_{AB}$ is \emph{locally broadcastable} (LB)
if there exist local maps $\Theta_{A}:A \rightarrow AA'$,  
$\Theta_{B}:B \rightarrow BB'$ such that 
$\sigma_{AA',BB'}\equiv \Theta_{A} \otimes \Theta_{B}(\rho_{AB})$
is a broadcast state for $\rho$.
\end{definition}

No entangled state is LB, as no entangled state can be broadcast even by LOCC (see Proposition 1 in \cite{PRLdong}). On the contrary, every CC state is LB by cloning locally its biorthonormal eigenbasis.
We provide now a necessary and sufficient condition for local broadcastability in terms of QMI.

\begin{theorem}
\label{thm:LB}
A state $\rho_{AB}$ is LB if and only if there exist a broadcast state $\sigma_{AA',BB'}$ for $\rho_{AB}$ such that $I(\rho_{A:B})=I(\sigma_{AA':BB'})$.
Moreover, any broadcast state $\sigma_{AA':BB'}$ satisfying the latter condition can be obtained from $\rho$ by means of local maps. 
\end{theorem}
\begin{proof}
If $\rho=\rho_{AB}$ is LB then there exist a broadcast state $\sigma=\sigma_{AA':BB'}\equiv (\Theta_{A} \otimes \Theta_{B})[\rho_{AB}]$. Since $\sigma$ is obtained from $\rho=\rho_{AB}$ by local operations, we have that $I(\sigma)\leq I(\rho)$, because local operations can not increase MI. Moreover, since $\sigma$ is a broadcast state, $\rho$ can be obtained by local operations from it, more precisely by local tracing. Indeed, $\rho=(\Tr_{A'}\ot\Tr_{B'})[\sigma]$, so that it must be $I(\sigma)\geq I(\rho)$. Therefore $I(\rho_{A:B})=I(\sigma_{AA':BB'})$. On the other hand,
let us now suppose there exist a broadcast state $\sigma$ for $\rho$ such that
$I(\rho_{A:B})=I(\sigma_{AA':BB'})$. We want to see it can be obtained by local broadcasting. Indeed, by taking $\Lambda_{AA'}=\Tr_{A'}$ and $\Lambda_{BB'}=\Tr_{B'}$, we have $I(\sigma)=I(\rho)=I((\Lambda_{AA'}\ot\Lambda_{BB'})[\sigma])$. By applying Lemma \ref{lem:petz}, we see there are local maps $\Theta_{A}=\Lambda^*_{AA'}$ and $\Theta_{B}=\Lambda^*_{BB'}$ that locally broadcast $\rho$ into $\sigma$.
\end{proof}

From Theorem \ref{thm:LB} we see that local broadcastability can be assessed by checking the existence of broadcast states with the same MI as the starting state.

We state now our second main result.
\begin{theorem}
\label{thm:ccbroadcast}
Classical-classical states are the only states that can be locally broadcast.
\end{theorem}
\begin{proof}
Given a LB state $\rho_{AB}$, consider any broadcast state $\sigma_{AA'BB'}$ satisfying $I(\rho)=I(\sigma)$, and let  measuring maps $\mathcal{M}$ and $\mathcal{N}$ be optimal for the sake of $I_{CC}(\rho)$.
Applying $\mathcal{M}$ and $\mathcal{N}$ on subsystems $A$ and $B$ of $\sigma$, we obtain:
$\tilde{\sigma}=(M_A\otimes N_B)[\sigma]=\sum_{ij} p_{ij} \proj{i_Aj_B}\ot\rho^{ij}_{A'B'}$.
Here, $p_{ij}=\tr{M_i^A\otimes N_j^B\ot \openone_{A'B'} \sigma}$ coincides with the optimal classical probability distribution for $\rho$, $\tr{M_i^A\otimes N_j^B \rho}$, because of the broadcasting condition, and $\rho^{ij}_{A'B'}=\Tr_{AB}(M_i^A\otimes N_i^B \sigma)/p_{ij}$. For the same reason, $\Tr_{AB}(\tilde{\sigma})=\Tr_{AB}(\sigma)=\sigma_{A'B'}=\rho_{AB}$. Thus, $I(\tilde{\sigma})=I(\rho)$, and at the same time
\begin{equation}
\label{eq:proof3}
\begin{split}
I(\tilde{\sigma})
								&=I(\{p_{ij}\})+\sum_ip^{A}_i S(\tau_{A'}^{i})+\sum_jp^{B}_j S(\tau_{B'}^{j})\\
								&-\sum_{ij}p_{ij}S(\rho^{ij}_{A'B'})\\
								&\geq I_{CC}(\rho)+\sum_{ij}p_{ij}I(\rho^{ij}_{A'B'}),
\end{split}
\end{equation}
where $p^A_i=\sum_{j}p_{ij}$,
$\tau_{A'}^{i}=\sum_j p_{ij}/p^{A}_i\rho_{A'}^{ij}$ (similarly for $p^B_i$ and $\tau_{B'}^{j}$). The inequality comes from the concavity of entropy: $\sum_i p^{A}_i S(\tau_{A'}^{i})\geq\sum_{ij}p_{ij}S(\rho_{A'}^{ij})$ (similarly for $B$), and we have used the fact that $I(\{p_{ij}\})=I_{CC}(\rho)$. Consider now any other measuring maps $\mathcal{\tilde{M}}$ and $\mathcal{\tilde{N}}$, and let them act on the (still quantum) systems $A'$ and $B'$ of $\tilde{\sigma}$, getting a state $\sigma^{CC}$. This corresponds simply to transforming each $\rho^{ij}_{A'B'}$ into some CC state $(\rho^{ij})^{CC}_{A'B'}(\{\tilde{M}_i\},\{\tilde{N}_j\})$. Thus, we have $I_{CC}(\sigma)\geq I(\sigma^{CC})=I_{CC}(\rho) + \sum_{ij}p_{ij}I\big((\rho^{ij})^{CC}_{A'B'}(\{\tilde{M}_i\},\{\tilde{N}_j\})\big)$,
for arbitrary $\{\tilde{M}_i\},\{\tilde{N}_j\}$, because the measurement maps $\mathcal{M}_A\otimes\mathcal{\tilde{M}}_{A'}$ and $N_B\otimes\tilde{N}_{B'}$ may not be the optimal ones to get $I_{CC}(\sigma)$. By the assumptions and by Theorem \ref{thm:LB}, $\sigma$ may be obtained from $\rho$ via local broadcasting, and by Observation \ref{obs:ICC} it must be $I_{CC}(\sigma)\leq I_{CC}(\rho)$. Therefore, we have $I_{CC}(\sigma)=I_{CC}(\rho)$. This means that $I((\rho^{ij})^{CC}_{A'B'}(\tilde{M},\tilde{N}))$ must be zero for any non vanishing $p_{ij}$. Choosing $\tilde{M},\tilde{N}$ repeatedly to be optimal for every $\rho^{ij}_{A'B'}$, one concludes that it must be $I_{CC}(\rho^{ij}_{A'B'})=0$ for every $i,j$ such that $p_{ij}>0$, so that, according to Observation \ref{obs:ICC}, it must be $\rho^{ij}_{A'B'}=\rho^{ij}_{A'}\ot\rho^{ij}_{B'}$. Moreover to have equality in \ref{eq:proof3}, it must be that $\rho^{ij}_{A'}=\rho^{i}_{A'}$ and $\rho^{ij}_{B'}=\rho^{j}_{B'}$, because of the strong concavity of entropy. Thus, we have found that actually $\tilde{\sigma}$ is a classical-classical state,
$\tilde{\sigma}=\sum_{ij} p_{ij} (\proj{i_A}\ot\rho^{i}_{A'})\ot(\proj{j_B}\ot\rho^{j}_{B'})$,
so that $I(\rho)=I(\sigma)=I_{CC}(\tilde{\sigma})=I_{CC}(\sigma)=I_{CC}(\rho)$, because of Observation \ref{obs:ICC}. Therefore, according to Theorem \ref{thm:cc}, $\rho$ is also classical-classical.
\end{proof}

One immediately realizes that the essential assumptions used to prove that $\rho_{AB}$ is CC are: (i) $\sigma_{AA'BB'}$ is obtained from $\rho$ by local maps; (ii) $I(\sigma_{AB})=I(\sigma_{A'B'})=I(\rho_{AB})$. Indeed, thanks to Lemma \ref{lem:petz}, these conditions  mean that $\rho_{AB},\sigma_{AA'BB'},\sigma_{AB},\sigma_{A'B'}$ are all connected by local maps. Thus, with slight changes in the proof of Theorem \ref{thm:ccbroadcast} one can obtain the following stronger result.

\begin{theorem}
\label{thm:no-broad-I}
Given a state $\rho_{AB}$, there exists a state $\sigma_{AA'BB'}$ with $I(\sigma_{AB})=I(\sigma_{A'B'})=I(\rho_{AB})$, that can be obtained from $\rho_{AB}$ by means of local operations, if and only if $\rho_{AB}$ is classical-classical.
\end{theorem}

The just stated result represents a no-broadcasting theorem, more precisely, a no-\emph{local}-broadcasting theorem, for correlations as measured by a single number, mutual information. Indeed, we do not require the (structure of the) state to be broadcast, rather is the \emph{amount of correlations} that counts. As such, the present result points out a fundamental difference between classical and quantum mutual information: correlations measured by the latter cannot be shared, in the broadcasting sense, as soon as the state can not be interpreted as describing the joint state of some classical registers. 
We remark that our result regards single states $\rho_{AB}$ of a bipartite system, while the standard no-broadcasting theorem~\cite{barnum-broad} refers to a set of two or more states $\{\rho_i^B\}$ of a single system B. The no-broadcasting theorem says that there is a single map $\Gamma:B\rightarrow AB$ such that $\Tr_A(\Gamma[\rho_i])=\Tr_B(\Gamma[\rho_i])=\rho_i^{B}$, if and only if the the states $\rho_i^B$ commute. Also this condition may be interpreted in terms of classicality of the states, in the following sense: when all the states are diagonal in the same basis, they may be considered distribution probabilities over possible classical states of the same classical register. We notice that our Theorem \ref{thm:ccbroadcast}, implies the standard no-broadcasting theorem. In order to see this, it is sufficient to consider a CQ state $\sigma=\sum_i p_i \proj{i}\ot\rho_i^B$, with $p_i>0$ for each $i$. Indeed, if states $\{\rho_i^B\}$ can be broadcast, then also $\sigma$ can be locally broadcast; our results say that $\sigma$ is LB if and only if it is $CC$, i.e. if and only if states $\rho_i^B$ commute.

All the previous results can be extended to the multipartite setting, by considering the following multipartite version of mutual information: $I(A_1:A_2:\ldots:A_n)
=S(\rho_{A_1A_2\ldots A_n}||\rho_{A_1}\ot\rho_{A_2}\ot \cdots \ot \rho_{A_n})$.
This quantity vanishes if and only if the state of the $n$ subsystems is completely factorized and does not increase under local operations.
All the other definitions are trivially extended to the multipartite case:
(i) a strictly classical correlated state is a probability multidistribution embedded in the quantum formalism;
(ii) given a state $\rho_{A_1A_2\ldots A_n}$, we say that $\tilde{\rho}_{A_1A_1'A_2A_2'\ldots A_nA_n'}$ is a broadcast state for $\rho$ if $\tilde{\rho}$ satisfies $\tilde{\rho}_{A_1A_2\ldots A_n}=\tilde{\rho}_{A'_1A'_2\ldots A'_n}=\rho_{A_1A_2\ldots A_n}$; (iii) a state can be made classical on chosen parties by local measuring maps; (iv) optimizing mutual information for the states obtained acting by measuring maps over an increasing number of parties, gives rise to a whole family of mutual information quantities. All Theorems remain valid, as Observation \ref{obs:ICC} and Lemma \ref{lem:petz} are immediately extended, while Lemma \ref{lem:holevo} generalizes to the case of a state that is classical with respect to all the parties but one. 

In conclusion, we characterized operationally the set of classical-classical states, i.e. states that correspond essentially to the description of correlated classical registers. We showed that they are the only states for which correlations, as measured by mutual information, can be totally transferred from the quantum to the classical world. Furthermore, they are the only states that can be locally broadcast. A even stronger result was derived in terms of mutual information alone, without imposing the broadcast condition for states: correlations, as quantified by such a scalar quantity, can be locally broadcast only for classical-classical states. Thus, our results show that also separable non-CC  states exhibit a certain degree of quantumness, and also lead to some natural ways to quantify the degree of non-classicality. E.g., one may consider the gap $\Delta_{CC}(\rho)=I(\rho)-I_{CC}(\rho)$, or, similarly to what done in~\cite{broad1}, the minimal difference $\Delta_b(\rho_{AB})=\min_{\sigma_{AA'BB'}}I(\sigma_{AA':BB'})-I(\rho_{AB})$, between the mutual information of a two-copy broadcast state $\sigma_{AA'BB'}$ and the mutual information of the state $\rho_{AB}$ itself. Theorems \ref{thm:cc}, and \ref{thm:LB} and \ref{thm:ccbroadcast}, respectively, make sure that such quantities are strictly positive for all non classical-classical states, and in particular entangled states. Actually, the gap $\Delta_{CC}$ resembles the \emph{discord} introduced in \cite{zurek}: the latter corresponds to the gap $I-I_{\tilde{C}Q}$, where $\tilde{C}$ means that the measuring map which gives rise to $I_{\tilde{C}Q}$ is chosen among complete projective measurements rather than POVMs, as in the case of $I_{CQ}$. A further analysis of the role of entanglement in the quantumness of correlations, as well of how our approach may lead to a non-trivial quantification of entanglement will appear somewhere else.

We thank G. Adesso, B. Kraus, M. Horodecki and C. E. Mora for discussions. Work supported by EC (through the IP SCALA) and the Austrian Science Fund (FWF).


\begin{thebibliography}{99}
\bibitem{footnote}This task is related to, but different from, the standard broadcasting task~\cite{barnum-broad}. For recent developments concerning the latter, see \cite{barnum}.
\end{thebibliography}
\end{document}